\documentclass{icrc2009}

\usepackage{graphicx}   
\usepackage[caption=false]{caption}    
\usepackage[font=footnotesize]{subfig} 
\usepackage{fixltx2e}
\usepackage{url}

\newcommand{\shorttitle}[1]%
{\markboth{Proceedings of the 31\MakeLowercase{$^{st}$} ICRC, {\L}\'{o}d\'{z} 2009}{#1} }
\newcommand{\etal}{\MakeLowercase{\textit{et al. }}} 

\begin{document}
\title{AGILE and blazar studies}

\author{\IEEEauthorblockN{M. Marisaldi\IEEEauthorrefmark{1}, F. D'Ammando\IEEEauthorrefmark{2}\IEEEauthorrefmark{3},
			  S. Vercellone\IEEEauthorrefmark{4},
                          I. Donnarumma\IEEEauthorrefmark{2},
                          A. Bulgarelli\IEEEauthorrefmark{1},
			  A. W. Chen\IEEEauthorrefmark{5}\IEEEauthorrefmark{6}\\
   A. Giuliani\IEEEauthorrefmark{5},
			  F. Longo\IEEEauthorrefmark{7},
			  L. Pacciani\IEEEauthorrefmark{2},
			  G. Pucella\IEEEauthorrefmark{2},
			  M. Tavani\IEEEauthorrefmark{2}\IEEEauthorrefmark{3} and
                          V. Vittorini
			  \IEEEauthorrefmark{2}\IEEEauthorrefmark{6} \\ (on
  behalf of the AGILE Team)}
                            \\
\IEEEauthorblockA{\IEEEauthorrefmark{1}INAF-IASF Bologna, Via Gobetti 101, 40129 Bologna, Italy}
\IEEEauthorblockA{\IEEEauthorrefmark{2}INAF-IASF Roma, Via Fosso del Cavaliere
			  100, 00133 Roma, Italy\\}
\IEEEauthorblockA{\IEEEauthorrefmark{3}Dip. di Fisica, Univ. ``Tor Vergata'', Via della Ricerca Scientifica 1, 00133
Roma, Italy\\}
\IEEEauthorblockA{\IEEEauthorrefmark{4}INAF-IASF Palermo, Via Ugo La Malfa 153, 90146 Palermo, Italy\\}
\IEEEauthorblockA{\IEEEauthorrefmark{5}INAF-IASF Milano, Via E. Bassini 15, 20133 Milano, Italy\\}
\IEEEauthorblockA{\IEEEauthorrefmark{6}CIFS-Torino, Viale Settimio Severo 3, 10133 Torino, Italy\\}
\IEEEauthorblockA{\IEEEauthorrefmark{7}Dip. di Fisica and INFN, Via Valerio 2, 34127 Trieste, Italy\\}
}

\shorttitle{M. Marisaldi \etal AGILE and blazar studies }
\maketitle

\begin{abstract} During the first two years of observation, AGILE detected
several blazars at high significance: 3C 279, 3C 454.3, PKS 1510--089, S5
0716+714, 3C 273, W Comae, Mrk 421 and PKS 0537--441. We obtained long-term coverage of 3C 454.3, for a total of more than
three months at energies above 100 MeV. 3C 273 was the first blazar detected simultaneously by the AGILE gamma-ray detector and by its
hard X-ray monitor. S5 0716+714, an intermediate BL Lac object,
exhibited a very fast and intense gamma-ray transient during an
optical high-state phase, challenging the current theoretical models for
energy extraction from a rotating black hole, while W Comae and Mkn 421 were detected during an AGILE Target of Opportunity (ToO) repointing, and
were simultaneously detected by Cherenkov telescopes.
Thanks to the rapid dissemination of our alerts, we were able to
obtain multi-wavelength ToO data from other observatories such as $Spitzer$,
$Swift$, RXTE, $Suzaku$, INTEGRAL, MAGIC, VERITAS, as well as radio-to-optical
coverage by means of the WEBT Consortium and REM, allowing
us to study truly simultaneous spectral energy distributions from
the radio to the TeV energy band.
  \end{abstract}

 
\section{Introduction}
Among active galactic nuclei (AGNs), blazars are a subclass characterized by the
emission of strong non-thermal radiation across the entire electromagnetic spectrum,
and in particular intense and variable $\gamma$-ray emission above 100 MeV
(see \cite{hart99}). The typical observational properties include
irregular, rapid and often very large variability, apparent super-luminal
motion in the radio emission of the jet, flat radio spectrum, high and variable polarization at radio and
optical frequencies. These features are interpreted as the result of the
emission of electromagnetic radiation from a relativistic jet that is viewed
closely aligned to the line of sight \cite{BR} and \cite{UP}. 

Blazars emit across several decades of energy, from radio
to TeV energy bands, and thus they are the perfect candidates for simultaneous
observations at different wavelengths. Multi-wavelength studies of variable
$\gamma$-ray blazars have been carried out 
since the beginning of the 1990s, thanks to the EGRET instrument onboard Compton Gamma-Ray Observatory
(CGRO), providing evidence that the Spectral Energy Distributions (SEDs) of
the blazars are typically double humped with the first peak occurring in the
IR/optical band in the so-called $red$ $blazars$ (including Flat Spectrum Radio
Quasars, FSRQs, and Low-energy peaked BL Lacs, LBLs)  and in UV/X-rays in the
so-called $blue$ $blazars$ (including High-energy peaked BL Lacs, HBLs). This first peak is
interpreted as synchrotron radiation from high-energy electrons in a
relativistic jet. The SED second component, peaking at MeV--GeV energies in
$red$ $blazars$ and at TeV energies in $blue$ $blazars$, is commonly interpreted as inverse Compton
scattering of seed photons by highly relativistic
electrons \cite{Ul}, although other models have been proposed
(see e.g. \cite{Bo} for a recent review). 

3C 279 is the best example of multi epoch studies at different frequencies
performed by EGRET
during the period 1991--2000 \cite{hart01}. Nevertheless, only a few object were detected on a timescales of about two
weeks in the $\gamma$-ray band, and simultaneously monitored at different
energies in order to obtain a wide multi-frequency coverage. 
Since its launch, the AGILE satellite \cite{Ta} detected several flaring blazars, and
thanks to the fast quick-look analysis procedure, extensive multi-wavelength
campaign were organized for many of them.
 \\
 

 \begin{figure*}[!t]
   \centerline{\subfloat{\includegraphics[width=2.7in]{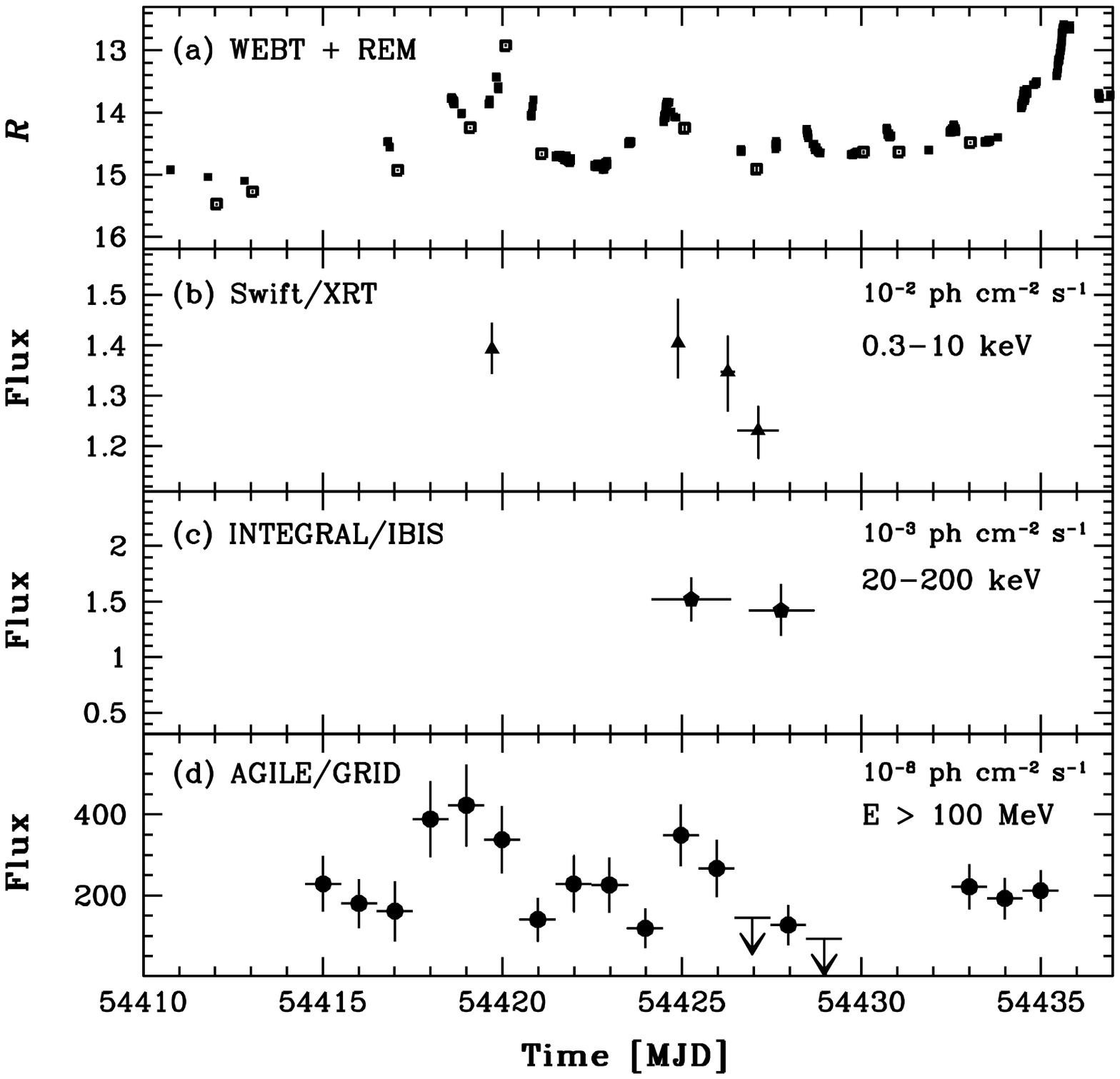} \label{sub_fig1}}
              \hfil
              \subfloat{\includegraphics[width=3.5in]{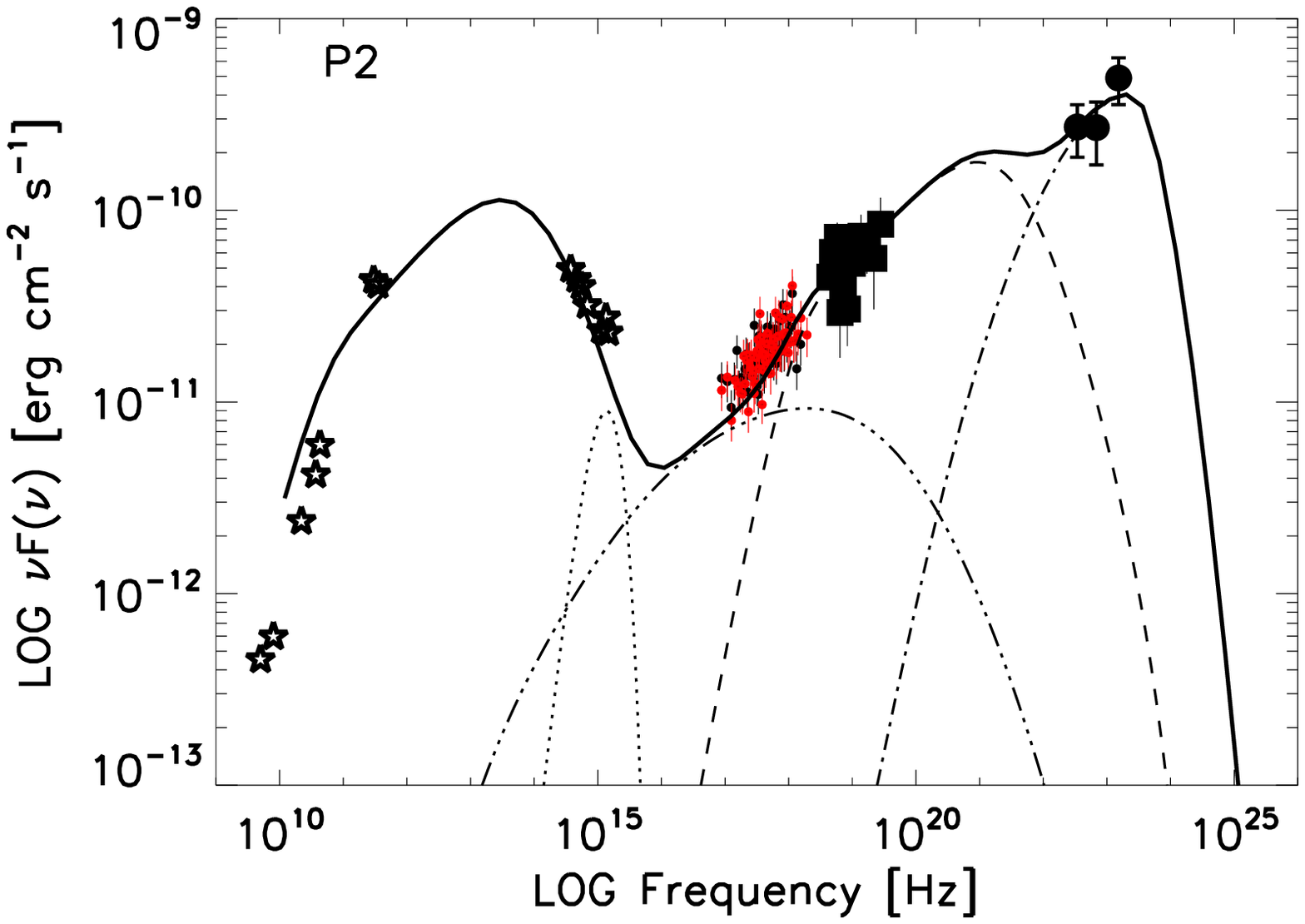} \label{sub_fig2}}
             }
   \caption{{\em Left panel:} Simultaneous light curves of 3C 454.3 acquired during the period 2007 November 6 - December 3. From bottom to the top AGILE/GRID, INTEGRAL/IBIS,
   $Swift$/XRT, REM and GASP-WEBT data. {\em Right panel:} Spectral Energy Distribution of 3C
   454.3 including AGILE/GRID, INTEGRAL/IBIS, $Swift$/XRT, $Swift$/UVOT and GASP-WEBT
   data collected in the period 2007 November 19--22. The dotted, dashed, dot-dashed, and triple-dot-dashed lines represent
   the accretion disk black body, the external Compton on the disk, the external Compton
   on the BLR and the SSC radiation, respectively.}
   \label{3C454_Nov07}
 \end{figure*}

\section{The AGILE Blazar sample}
  
AGILE is an Italian Space Agency (ASI) mission successfully launched on 23
April 2007 and capable of observing cosmic sources simultaneously in X-ray
(18--60 keV) and gamma-ray (30 MeV--30 GeV) bands with the coded-mask hard X-ray
imager (SuperAGILE) and the Gamma-Ray Imaging Detector (GRID),
respectively. Gamma-ray observations of blazars are a key scientific project
of the AGILE satellite and in the last two years AGILE detected several
blazars during high $\gamma$-ray emission. However, up to now a few sources were
detected more than once during very high activity state: S5 0716+714, PKS
1510--089 and 3C 454.3. We note that we detected
at least one object for each blazar category: 3C 454.3 (FSRQ), PKS 0537--441
(LBL), S5 0716+714 (intermediate BL Lac, IBL) and Mrk 421 (HBL). Moreover,
the $\gamma$-ray activity timescales goes from a few days (e.g. S5 0716+714)
to several weeks (e.g. PKS 1510--089 and 3C 454.3) and the flux variability for E $>$ 100 MeV
could be negligible (e.g. 3C 279) or extremely high (e.g. 3C 454.3 and PKS 1510--089). In the
following sections we will present the most interesting results on
multi-wavelength observations of some of the sources detected by AGILE.

\section{3C 454.3}

 3C 454.3 is the blazar which exhibited the most variable activity in the
 $\gamma$-ray sky in the last two years. In the period July 2007--January 2009
 the AGILE satellite monitored intensively 3C 454.3 together with $Spitzer$,
 GASP-WEBT, REM, MITSuME, $Swift$, RXTE, $Suzaku$ and INTEGRAL observatories,
 yielding the longest multi-wavelength coverage of this $\gamma$-ray quasar so
 far \cite{Ver1y}. It is clear that this source
 is playing the same role for AGILE as 3C 279 had for EGRET. 

\begin{figure}[!b]
  \centering
  \includegraphics[width=3.0in]{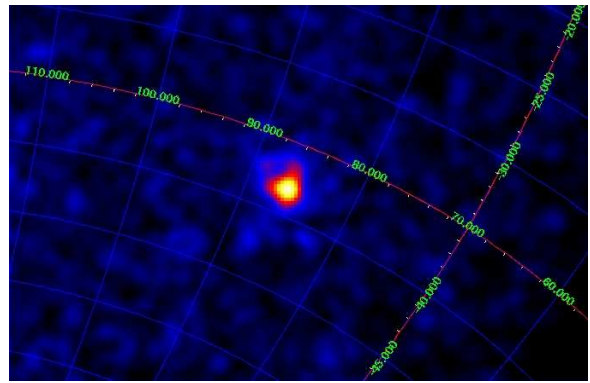}
  \caption{Gaussian smoothed couts map for E $>$ 100 MeV in Galactic coordinates integrated over
  the AGILE oberving period 24--30 July 2007, showing the detection of 3C
  454.3 during flaring activity.}
 \label{SED_3C454_Dec}
\end{figure}

AGILE detected 3C
 454.3 for the first time during a dedicated Target of Opportunity (ToO)
 activated immediately after an extremely bright optical flare in mid-July
 2007 (see Fig. 2) \cite{Ver08}. During a 6-day observation, the average $\gamma$-ray flux was
 $F_{E>100 MeV}$ = (280 $\pm$ 40) $\times$ 10$^{-8}$ photons cm$^{-2}$
 s$^{-1}$, more than a factor two higher than the maximum value reported by
 EGRET. Moreover, the peak flux on a daily timescale was of the order of (400
 $\pm$ 100) $\times$ 10$^{-8}$ photons cm$^{-2}$ s$^{-1}$. Since this
 detection, this source become the most luminous object in the AGILE sky.

Subsequently, a multi-wavelength campaign on 3C 454.3 was organized during
 November 2007 as reported in \cite{Ver09}. We monitored the source for about 18 days, obtaining
 simultaneous observations also from radio to hard X-ray energy bands. Figure 1
 (left panel) shows the optical (WEBT and REM), soft X-rays
 ($Swift$/XRT), hard X-rays (INTEGRAL/IBIS) and $\gamma$-rays (AGILE/GRID) light
 curves. The $\gamma$-ray flux appears to be variable on a short timescale
 (24-48 hours), while the soft and hard X-ray energy bands show a less
 pronunced variability. The average $\gamma$-ray flux over the entire campaign
 is $F_{E>100 MeV}$ = (170 $\pm$ 13) $\times$ 10$^{-8}$ photons cm$^{-2}$
 s$^{-1}$. The $R$-band optical flux appears extremely variable, with a
 brigthening of several tenths of magnitude in a few hours. A correlation
 analysis between optical and $\gamma$-ray flux variations is consistent with
 no time-lags. The average $\gamma$-ray photon index during the entire
 campaign is substantially harder than that reported in the past by EGRET
 ($\Gamma_{AGILE}$ = 1.73 $\pm$ 0.16 vs. $\Gamma_{EGRET}$ = 2.22 $\pm$
 0.06). Figure 1 (right panel) shows the SED, in which the dominant emission
 mechanism over 100 MeV seems to be the inverse Compton scattering of
 relativistic electrons in the jet on the external photons from the Broad Line
 Region (BLR).

 In December 2007, the activity of the source continued to be high and AGILE collected
 other 15 days of monitoring on 3C 454.3, detecting the source with a flux of
 the order of 250 $\times$ 10$^{-8}$ photons cm$^{-2}$ s$^{-1}$. Therefore, we triggered
 a second multi-wavelength campaign with $Spitzer$, REM, WEBT, MITSuME, $Swift$ and
 $Suzaku$. The hard $\gamma$-ray spectrum and the high Compton dominance
 observed by AGILE suggested the contribution of external Compton of seed photons from a hot
 corona. Results of this campaign will appear in  \cite{Don09}. During
 May--June 2008, AGILE performed the most intensive and long campaign on 3C 454.3, resulting
 in an almost uninterrupted 50-day long monitoring, collecting also data with WEBT, REM, $Swift$ and RXTE. This long observation
 showed the highly variable nature of 3C 454.3, both on short and long time
 scales. The multi-wavelength study of the source will be presented in a
 forthcoming paper \cite{Ver1y}. 

\begin{figure}[!b]
  \centering
  \includegraphics[width=3.25in]{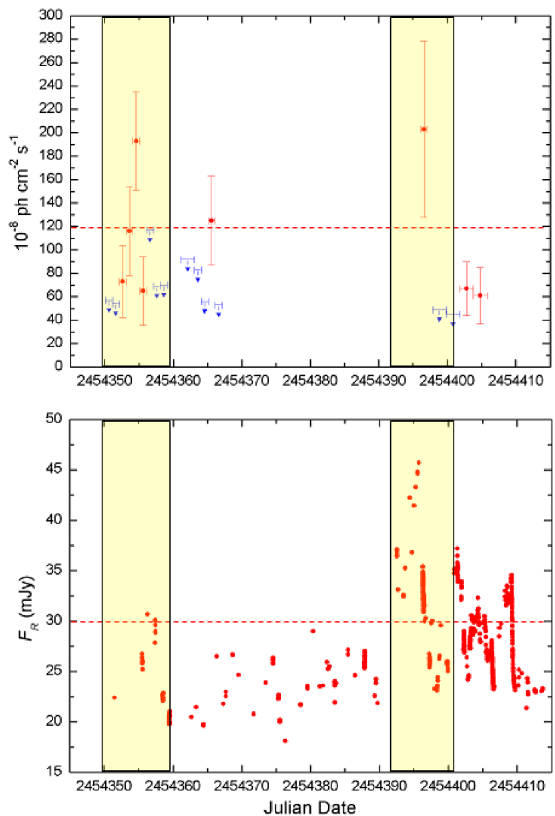}
  \caption{In the top panel, the AGILE/GRID light curve with 1-day or 2-day
  resolution for fluxes in units of 10$^{-8}$ ph cm$^{-2}$ s$^{-1}$ for E
  $>$ 100 MeV. The downward arrows represent 2-$\sigma$ upper limits. In the
  bottom panel, the $R$-band optical light curve as observed by GASP-WEBT. The
  shaded regions indicate the two high-activity periods in $\gamma$-rays.}
  \label{0716_lc}
 \end{figure}

\section{S5 0716+714}
 
\begin{figure}[!t]
  \centering
  \includegraphics[width=3.15in]{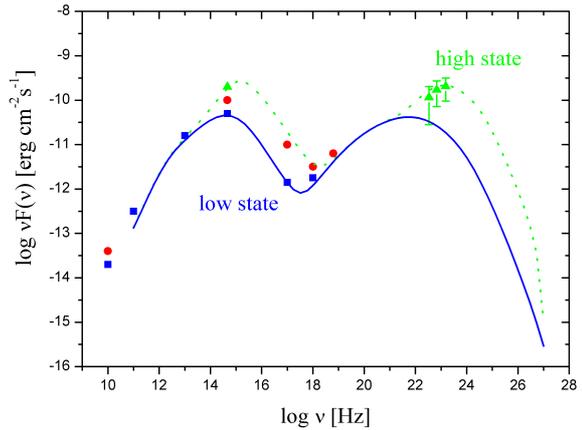}
  \caption{SED of S5 0716+714, including GASP-WEBT optical data
  quasi-simultaneous with the AGILE/GRID $\gamma$-ray observation in
  September 2007 (green triangles). Historical data over the electromagnetic spectrum relative to a
  ground state of the source and EGRET non-simultaneous data are
  represented with blue squares. Red dots represent historical data simultaneous with a high
  X-ray state.}
  \label{0716_SED}
 \end{figure}

The intermediate BL Lac object S5 0716+714 was observed by AGILE during two
different periods: 4--23 September and 23 October -- 1 November 2007, as
discussed in \cite{Chen08}. In mid-September the source showed an high $\gamma$-ray activity with an average
flux of F$_{E>100 MeV}$ = (97 $\pm$ 15)$\times$10$^{-8}$ photons cm$^{-2}$ s$^{-1}$
and a peak flux of F$_{E>100 MeV}$ =
(193 $\pm$ 42)$\times$10$^{-8}$ photons cm$^{-2}$ s$^{-1}$. The flaring activity
appeared to be very fast with an increase of the $\gamma$-ray flux by a factor
of four in three days (see Fig. 3). This source was detected by EGRET in a low/intermediate
$\gamma$-ray level (F$_{E>100 MeV}$ = (20--40) $\times$10$^{-8}$ photons cm$^{-2}$
s$^{-1}$), therefore the flux detected by AGILE is the highest flux detected by this object and one of the most high flux observed by
a BL Lac object. An almost simultaneous GASP-WEBT optical campaign started
after the AGILE detection and the resulting SED is consistent with a
two-components synchrotron self Compton (SSC) model (Fig. 4). Recently \cite{Ni} has estimated the
redshift of the source (z = 0.31 $\pm$ 0.08) and this allowed us to calculate the total power
transported in the jet, which results extremely high and at limit of the maximum
power generated by a spinning black hole of 10$^{9}$ M$_\odot$ \cite{Vittorini}. 

During October 2007, AGILE detected the source for the second time, as
reported in \cite{Giommi}, at a flux about a factor 2 lower than
September one with no significant variability. Instead, simultaneously $Swift$ observed strong
variability in soft X-ray, moderate variability at optical/UV and
approximately constant hard X-ray flux. Also this behaviour is compatible with
the presence of two different SSC components in the SED.

\section{3C 273}

3C 273 was the first extragalactic source detected simultaneously by the GRID
and SuperAGILE detectors during a pre-planned campaign over three
weeks between mid-December 2007 and January 2008 with also simultaneous REM,
$Swift$, RXTE and INTEGRAL coverage. During this campaign, whose results are
reported in detail in \cite{Pacciani}, the average flux in the 20--60~keV energy
band is (23.9 $\pm$ 1.2) mCrab, whereas the source was detected by the GRID only in the second week, with an average flux of F$_{E>100 MeV}$ =
(33 $\pm$ 11)$\times$10$^{-8}$ photons cm$^{-2}$ s$^{-1}$. The comparison of
the light curves seems to indicate no optical variability during the whole
campaign and a possible anti-correlation between the
$\gamma$-ray emission and the soft and hard X-rays. The SED is consistent with
a leptonic model where the soft X-ray emission is produced by the combination
of SSC and EC models, while the hard X-ray and $\gamma$-ray emission is due to
external Compton scattering by thermal photons of the disk. The spectral variability
between the first and the second week is consistent with an acceleration
episode of the electron population responsible for the synchrotron emission,
leading to a shift in the inverse Compton peak toward higher energies (see
Fig. 5).

\begin{figure}[!t]
  \centering
  \includegraphics[width=3.25in]{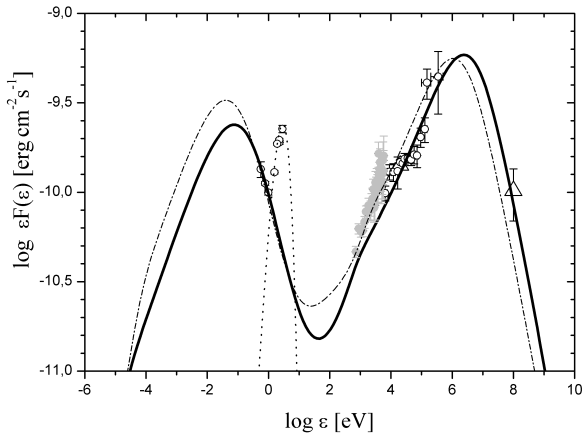}
  \caption{Spectral Energy Distribution of 3C 273 for the second week of
  observation including AGILE/GRID data (triangle), INTEGRAL/ISGRI and JEM-X data
  (open circles), $Swift$/UVOT and REM data. The grey data refers to $Swift$/XRT
  observations, performed in the third week. The model of the first week is
  reported as a dot-dashed line for comparison.}
  \label{3C273_SED}
 \end{figure}

\section{TeV Blazars: W Comae and Mrk 421}

On 2008 June 8, VERITAS announced the detection of a TeV flare
from W Comae \cite{Swordy}. About 24 hours later, AGILE was already pointing towards the
source and detected it \cite{Verrecchia}. This source belongs to an AGILE
Cycle-1 Guest Investigator (E. Pian), and the results of a multi-wavelength
campaign involving $Swift$, AGILE and VERITAS will appear in a forthcoming paper, see
\cite{Maier} in these proceedings for more details. 

During the ToO observation towards W Comae, AGILE detected also
the HBL object Mrk 421 both with SuperAGILE (see \cite{Costa}) and GRID (preliminary results are reported
in \cite{Pittori}). A multi-wavelength campaign with simultaneous MeV--GeV
(AGILE) and TeV (MAGIC and VERITAS) data, together with WEBT, RXTE and $Swift$
observations was organized and the results are presented in \cite{Donnarumma}
and discussed also in these proceedings \cite{Wagner}. 

\section{Conclusions}

AGILE, during its first two years of sky monitoring, demonstrated the
importance of its wide field of view ($\sim$2.5 sr) in detecting transient
sources at high off-axis angle. Moreover, its unique combination of a
$\gamma$-ray detector with an hard X-ray detector allowed us to study in
detail the high energy part of the blazar SEDs. The synergy between the AGILE
wide field of view, its fast response to external triggers and the
availability of network of ground-based telescope, allowed us to obtain a multi-wavelength coverage for almost all the detected sources. Up to now, AGILE detected already
known $\gamma$-ray emitting blazars. This evidence, combined with the early
$Fermi$-LAT results, which show a predominant fraction of already known
$\gamma$-ray emitters among its flaring sources, poses possible constraints on
the emitting properties of the blazars. Future AGILE and $Fermi$-LAT observations
will be fundamental to unveil the real nature of the blazars.

\section*{Acknowledgments}

The AGILE Mission is funded by the Italian Space Agency (ASI) with scientific and programmatic participation by the
Italian Institute of Astrophysics (INAF) and the Italian Institute of Nuclear Physics (INFN).



\end{document}